\documentclass{article}
\usepackage{amsfonts}
\usepackage{amsmath}
\usepackage[pdftex]{graphicx}

\begin{document}

\title{Defects in
Four-Dimensional Continua: a Paradigm for the Expansion of the Universe?}
\author{Angelo Tartaglia\\
Dipartimento di Fisica, Politecnico di Torino, and INFN Torino\\
Corso Duca degli Abruzzi 24, I-10129 Torino, Italy\\[4pt]}

\maketitle

\begin{abstract}
The presence of defects in material continua is known to produce internal
permanent strained states. Extending the theory of defects to four
dimensions and allowing for the appropriate signature, it is possible to
apply these concepts to space-time. In this case a defect would induce a
non-trivial metric tensor, which can be interpreted as a gravitational
field. The image of a defect in space-time can be applied to the description
of the Big Bang. A review of the four-dimensional generalisation of defects
and an application to the expansion of the universe will be presented.
\end{abstract}

\section{Introduction}

\label{sec:1}

The correspondence between the space-time description typical of the general
relativity theory (GR) and the geometrical properties of continua has remote
roots in the ether theories of the XIXth century (see some interesting
references to 1839 Mac Cullagh theory in a review by A. Unzicker \cite%
{unzicker}). More specifically a formal link between moving dislocations and
special relativity was pointed out by Frank in 1949 \cite{frank}, then
variously discussed by a number of other authors (cited in sect. 2.1 of \cite%
{unzicker}). It is the very geometrization of space-time which immediately
suggests a correspondence with material continua, their metric properties,
and the theory of elasticity. This longly known analogy has been, and is now
and then, revived, but has never been taken too seriously and/or used as a
constitutive theory of space-time. There are of course philosophical reasons
for this mistrust, in a description of our universe basically dualistic
(space-time on one side, matter/energy on the other), where the attribute of
"reality", whatever it is, is easily assigned to matter/energy and rather
ambiguously recognized for space-time. Even within the framework of
relativity it is in practice hardly accepted the idea that time (apart from
signature) is really like the other dimensions of space and that space-time
with its metric properties is something more than a simple conceptual
artifact. In the present work I shall review the essentials of an
interpretation of space-time as a real, though peculiar, continuum where
defects play a fundamental role and I will draw some cosmological
conclusions from this approach. I am in fact trying to be consistent and to
seriously take space-time as an existing real entity, most in the sense of
what Einstein said in an address delivered on May 5th, 1920, in the
University of Leyden: "\ldots\ \emph{according to the general theory of
relativity space is endowed with physical qualities; in this sense,
therefore, there exists an ether. ... But this ether may not be thought of
as endowed with the quality characteristic of ponderable media, as
consisting of parts which may be tracked through time. .}.." \cite{einstein}.

There is a specific motivation to try and explore again the present
description of space-time. Since nine years or so the observation of cosmic
phenomena has evidenced what has reasonably been interpreted as an
accelerated expansion of the universe. This behaviour has been initially
recognized considering the apparent magnitude of type Ia supernovae \cite%
{magnitude}\cite{super2} (SnIa). SnIa's are a special type of supernovae
which are commonly thought to be originated from white dwarfs in binary
systems, with an implosion mechanism based on the reaching of the
Chandrasekhar mass limit \cite{niemayer}; this mechanism leads to a more or
less fixed absolute luminosity which makes SnIa's allegedly good standard
candles \cite{candle}\cite{SnIa}. The fact that the observed luminosity of
such supernovae appears to be systematically smaller than what expected from
their cosmic redshift, suggests the idea of an accelerated expansion.

The discovery of the acceleration has stimulated an intense and vast
theoretical effort in order to explain the unexpected behaviour. In general
the attempts of finding the reasons for the acceleration are based on some
mechanism able to produce a negative pressure on space-time, which is mostly
attributed to some "dark" (i.e. otherwise unseen) energy component present
in the universe. This dark energy ranges from the simplest (and most
effective) cosmological constant, uselessly introduced by Einstein in order
to avoid the whole expansion of the universe, to more sophisticated variants
of some exotic energy fluid endowed with a non-standard equation of state.
Other attempts, instead of directly introducing dark actors, concentrate in
looking for heuristic forms of the space-time Lagrangian, other than the
standard Einstein-Hilbert one.

The approach which is outlined here tries rather to build on the analogy
with known, even though enlarged and extended, properties of continua. As we
shall see the results are interesting and promising.

\section{Elasticity in N dimensions}

\label{sec:2}

Suppose you have a featureless material continuum. Perfect homogeneity is
assumed. In the view of a physicist, and assuming by default that all
appropriate mathematical conditions are fulfilled, it is natural to
associate with the continuum an Euclidean appropriately dimensioned manifold
with the related geometry. Each point in the continuum will naturally be
labelled by Cartesian coordinates (or any other sound coordinate system). If
now, in the case of a real continuum, we consider a boundary and apply to it
a set of forces globally in equilibrium, what happens is that our manifold
will be stretched, or, in other terms, each point within the chosen boundary
will be moved away from its original position. If we remove the applied
forces, we expect the induced strain to be nullified bringing each point
back to its former rest position. We know this is essentially a simple
pictorial description of elasticity. In terms of coordinates of a given
point, labelling the unstrained situation by means of $\xi $'s and the
strained one by means of $x$'s, we may write:%
\begin{equation}
x^{\mu }=\xi ^{\mu }+u^{\mu }  \label{uno}
\end{equation}%
ranging the $\mu $ apex from $1$ to $N$ (number of dimensions in the
manifold). The $u^{\mu }$'s are the components of the \emph{displacement
N-vector }leading from the original unstrained to the final strained
position. In practice the elastic deformation is described by giving a
peculiar displacement vector field. The displacement field will in general
not be uniform, otherwise we would have a rigid translation neutralizable by
a simple coordinate transformation. $u^{\mu }$'s may be expressed equally
well in terms of the $x$'s (intrinsic coordinates) or of the $\xi $'s
(extrinsic coordinates).

So far we may also write that $x$'s are differentiable functions of $\xi $'s
or that:%
\begin{equation}
dx^{\mu }=\frac{\partial x^{\mu }}{\partial \xi ^{\nu }}d\xi ^{\nu }
\label{diff}
\end{equation}

In practice we may compare two distinct manifolds, the unstrained or
reference one and the strained or natural one \cite{nat}, whose points are
in one to one correspondence. Comparing corresponding lengths in the two
manifolds leads to the definition of the strain tensor \cite{landau}

\begin{equation}
\varepsilon _{\mu \nu }=\frac{1}{2}\left( \frac{\partial u_{\mu }}{\partial
\xi ^{\nu }}+\frac{\partial u_{\nu }}{\partial \xi ^{\mu }}+\eta _{\alpha
\beta }\frac{\partial u^{\alpha }}{\partial \xi ^{\mu }}\frac{\partial
u^{\beta }}{\partial \xi ^{\nu }}\right)  \label{strain}
\end{equation}%
which enters the metric tensor of the strained manifold expressed in
extrinsic coordinates:%
\begin{equation}
g_{\mu \nu }=\eta _{\mu \nu }+2\varepsilon _{\mu \nu }  \label{metric}
\end{equation}

In eq.s (\ref{strain}) and (\ref{metric}) $\eta _{\mu \nu }$ is a component
of the metric tensor of the reference manifold. As previously written we
should expect this to correspond to an Euclidean geometry, however, in order
to apply our approach to space-time, we shall allow for a Minkowski
geometry. Actually the problem of the origin of the Lorentzian signature
from a fully symmetric $N$-dimesional manifold is an open one. We know that
we may formally go from the Euclidean to the Minkowskian geometry
introducing an imaginary coordinate that will act as time (Wick's rotation),
however no physically based mechanism for that has been found until now.

According to the assumptions made so far and to eq. (\ref{diff}) we expect
also:%
\begin{equation}
\eta _{\mu \nu }=g_{\alpha \beta }\frac{\partial x^{\alpha }}{\partial \xi
^{\mu }}\frac{\partial x^{\beta }}{\partial \xi ^{\nu }}  \label{trasfor}
\end{equation}

De SaintVenant's integrability condition for eq. (\ref{trasfor}) is:%
\begin{equation}
R_{\alpha \beta \mu \nu }\equiv 0  \label{desaint}
\end{equation}%
being $R_{\alpha \beta \mu \nu }$ a generic element of the Riemann tensor
for the natural manifold.

When condition (\ref{desaint}) is satisfied, globally defined $x^{\mu
}\left( \xi ^{1},\xi ^{2},...\right) $'s exist and in the same time the
strain of the natural manifold cannot be perceived from inside, or, which is
the same, the metric in intrinsic coordinates always turns to be globally
Euclidean (Minkowskian).

All this is typical of pure elasticity and the corresponding deformations
are not seen from within the deformed medium.

\section{Defects}

\label{sec:3}

The scenario outlined in the previous section becomes richer when the notion
of \emph{defect }is introduced. A defect may be reduced to its essentials
generalizing eq. (\ref{uno}) to the case of a singular displacement field.
Differently stated, we may substitute eq. (\ref{diff}) with

\begin{equation*}
dx^{\mu }=\Phi _{\nu }^{\mu }d\xi ^{\nu }
\end{equation*}%
where now $\Phi _{\nu }^{\mu }d\xi ^{\nu }$ is a non-integrable
vector-valued one-form.

\begin{figure}\label{fig:dislo}
\centerline{\includegraphics[width=10cm]{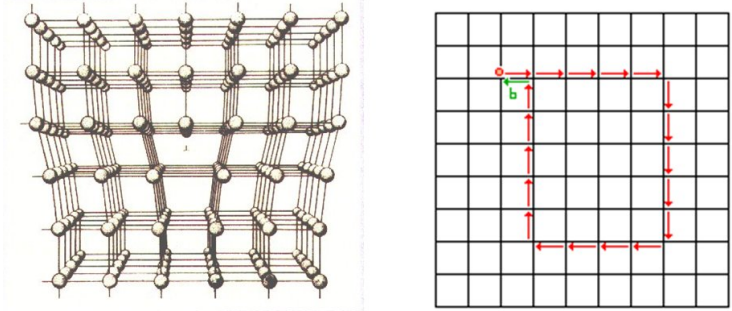}}
\caption{Schematic representation of a
dislocation. The pictorial view on the left is apparently applied to a
crystal, but this is absolutely not needed. On the right a closed path in
the real manifold is transposed to the reference manifold evidencing the
Burgers vector.}
\end{figure}

A complete classification of defects exists according to the peculiarities
of $\Phi _{\nu }^{\mu }$. Consider for instance a closed path in the
manifold such that%
\begin{equation}
\oint \Phi _{\nu }^{\mu }dx^{\nu }=-b^{\mu }\neq 0  \label{burgers}
\end{equation}%
The quantity $b^{\mu }$ produced by the integration in (\ref{burgers}) is a
component of the Burgers $N$-vector which expresses the fact that a closed
contour in the reference manifold does not correspond to a closed one in the
natural manifold, and vice-versa. Burgers vector measures the size of the
non-closure. If the defect which is the origin of this behaviour is thought
to be localized in the manifold, then we are considering a linear (or edge)
defect whose direction is given by the Burgers vector. Fig. (\ref{fig:dislo})
shows a typical dislocation in a crystal. Using a lattice makes the graph
clearer but is not necessary; a closed contour encircling the edge of the
singularity corresponds to an open path in the reference manifold, as it can
be seen on the right, where also the Burgers vector is drawn.
The integral in (\ref{burgers}) may be transformed by means of Stoke's
theorem, becoming%
\begin{equation*}
b^{\mu }=-\int\int \mathfrak{T}_{\alpha \beta }^{\mu }dx^{\beta }\wedge
dx^{\alpha }
\end{equation*}%
Now the integration is over the oriented surface enclosed in the former
integration path. $\mathfrak{T}_{\alpha \beta }^{\mu }$ is a dislocation
density and corresponds (being antisymmetric in $\alpha $ and $\beta $) to
the torsion tensor of the manifold.

Another well known type of edge defect is obtained when condition (\ref%
{desaint}) is not fulfilled. In this case parallelly transporting a vector $%
n $ along a closed contour ends with a rotated vector with respect to the
initial one: we have a disclination. It is indeed%
\begin{equation*}
\delta n^{\nu }=\oint dn^{\nu }=-\int\int R_{\mu \alpha \beta }^{\nu }n^{\mu
}dx^{\alpha }\wedge dx^{\beta }\neq 0
\end{equation*}%
The curvature tensor is now interpreted as a disclination density.

When considering space-time we have a four-dimensional manifold with
Lorentzian signature. What has been written concerning defects still holds
and it is remarkable that curvature (then gravity) can be read as a
consequence of the presence of defects in the manifold. In the case of
space-time edge defects can be qualified in terms of the Poincar\'{e} group.
In fact a general deformation of the continuum may be thought as a
combination of a translation and a local rotation; if $r$ is the $N$-vector
localizing a point in a given manifold and a given reference frame, the new
position after the deformation has been applied may be written \cite%
{puntigam}:

\begin{equation*}
r^{\prime }=T\left( r\right) +\Lambda \left( r\right) r
\end{equation*}

$T\left( r\right) $ and $\Lambda \left( r\right) $ respectively correspond
to local translation and Lorentz transformation operators. Within this
approach the presence of a defect is expressed in terms of the soldering one
form, which introduces the singular behaviour of the displacement field in
the typical line element (then the metric tensor):%
\begin{eqnarray*}
\omega &=&dx+\Gamma ^{T}+\Gamma ^{L}x \\
ds^{2} &=&\eta _{\mu \nu }\omega ^{\mu }\otimes \omega ^{\nu }
\end{eqnarray*}%
$\Gamma ^{T}$ and $\Gamma ^{L}$ represent respectively the translation and
the Lorentz connection. By this method 10 separate types of edge defects of
space-time are found \cite{puntigam}.

In order to complete the analogy between continuous media and Riemannian
manifolds we may recall that, at least in the linear elasticity theory,
there is a rather simple proportionality law between strains and stresses,
which is the general form of Hooke's law:
\begin{equation}
\sigma ^{\mu \nu }=C_{\alpha \beta }^{\mu \nu }\varepsilon ^{\alpha \beta }
\label{hooke}
\end{equation}%
where $C_{\mu \nu \alpha \beta }$ is the elastic modulus tensor, peculiar to
any given material continuum.

We may think to generalize (\ref{hooke}) to any number of dimensions, and,
even more, to space-time, although in this case the interpretation of the
stress tensor $\sigma $ is not at all obvious. This generalization may be
useful when looking for appropriate Lagrangians describing the state of a
given manifold, with or without defects. By the way, in an isotropic medium
(which could be the case of space-time) the elastic modulus tensor assumes
the simple form%
\begin{equation*}
C_{\alpha \beta \gamma \delta }=\lambda \eta _{\alpha \beta }\eta _{\gamma
\delta }+\mu \left( \eta _{\alpha \gamma }\eta _{\beta \delta }+\eta
_{\alpha \delta }\eta _{\beta \gamma }\right)
\end{equation*}%
depending on two parameters only: the Lam\'{e} coefficients $\lambda $ and $%
\mu $. The Hooke's law becomes:%
\begin{equation}
\sigma ^{\mu \nu }=\lambda \eta ^{\mu \nu }\varepsilon +2\mu \varepsilon
^{\mu \nu }  \label{omogeneo}
\end{equation}%
where $\varepsilon =\varepsilon _{\alpha }^{\alpha }$ is the trace of the
strain tensor.

\subsection{A point Defect}

\label{sec:3.1}

Edge defects are not the only possibility. A simple and interesting case is
the one of a point defect, which is graphically schematized on fig. (\ref%
{fig:punto}). The imagined process goes back to Vito Volterra \cite{volterra}%
, who studied plastic deformations and defects at the beginning of the XX
century. We may think of digging out of a continuum a sphere of the material
(whatever it is), then close the hole left behind, by pulling inwardly on
the walls.

\begin{figure}\label{fig:punto}
\centerline{\includegraphics[width=8cm]{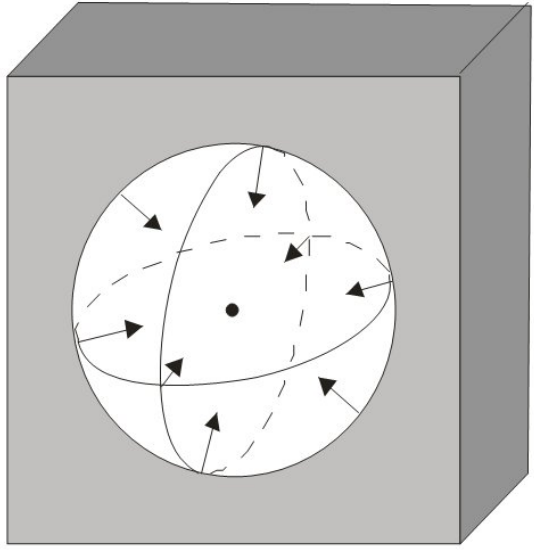}}
\caption{A point defect obtained by
contraction towards the center of an initial hollow sphere.}
\end{figure}

In order to avoid the problems of singularities, Volterra applied his ideal
process outside a fixed reference surface in the medium surrounding the
defect, but we may think to go on until the center. Of course we will
produce a singularity (the actual point defect) and induce everywhere a
spherically symmetric strained state. The displacement vector field is
easily written:%
\begin{equation}
u=\left( \psi \left( r\right) ,0,0,...\right) .  \label{radiale}
\end{equation}%
The only non-zero component is of course the radial one and it will depend
on the distance $r$ from the center (the defect) only.

From (\ref{radiale}) it is also easy to find the induced strain tensor. Let
us specialize to a four-dimensional manifold. The "radial" coordinate, with
Lorentzian signature, is indeed time ($\tau $, measured in meters); let us
then use polar coordinates (arbitrary origin for three-space) for the space
submanifold. The non-zero components of the strain tensor will be:
\begin{equation}
\begin{split}
\varepsilon _{00}& =\frac{1}{2}\left[ 2\frac{d\psi }{d\tau }+\left( \frac{%
d\psi }{d\tau }\right) ^{2}\right] \\
\varepsilon _{rr}& =\frac{\psi ^{2}}{2} \\
\varepsilon _{\theta \theta }& =\frac{\psi ^{2}}{2}r^{2} \\
\varepsilon _{\phi \phi }& =\frac{\psi ^{2}}{2}r^{2}\sin ^{2}\theta.
\end{split}
\label{centrale}
\end{equation}

Using eq. (\ref{metric}) we are now able to write the typical line element
in the strained manifold:%
\begin{equation*}
ds^{2}=\left( 1+\frac{d\psi }{d\tau }\right) ^{2}d\tau ^{2}-\left( 1-\psi
^{2}\right) \left( dr^{2}+r^{2}d\theta ^{2}+r^{2}\sin ^{2}\theta d\phi
^{2}\right) .
\end{equation*}

The presence of the defect is expressed by the discontinuity of the
derivative of $\psi $ in the origin, whereas $\psi \left( 0\right) $ is
finite and measures the "size" of the defect. Excluding the origin, it is
possible to redefine time choosing%
\begin{equation*}
dt=\left( 1+\frac{d\psi }{d\tau }\right) d\tau ,
\end{equation*}%
which gives%
\begin{equation*}
t=\tau +\psi +T_{0}.
\end{equation*}

In the origin ($t=\tau =0$) it is
\begin{equation*}
\psi \left( 0\right) =-T_{0}.
\end{equation*}

This change in the time coordinate, transforms the line element into:%
\begin{equation}
ds^{2}=dt^{2}-a^{2}\left( t\right) \left( dr^{2}+r^{2}d\theta ^{2}+r^{2}\sin
^{2}\theta d\phi ^{2}\right) ,  \label{RW}
\end{equation}%
with
\begin{equation}
a^{2}\left( t\right) =1-\psi ^{2}\left( t\right) .  \label{apsi}
\end{equation}

In (\ref{RW}) we immediately recognize a Robertson-Walker (RW) line element,
which is not a surprise since RW's is the most general line element for the
assumed symmetry. However here we have established a correspondence between
the RW metric and the presence of a defect in the origin, treating the
manifold as a material continuum.

\section{A Lagrangian For Space-Time}

\label{sec:4}

The correspondence found in the previous section between the metric of a
continuum with a point defect and the one of a RW universe is suggestive,
however in order to treat a universe with matter an appropriate Lagrangian
is needed. As it is well known, there is no general recipe for building
Lagrangians, so we may proceed in a more or less formal way or try to look
for analogies with already known situations.

\begin{figure}\label{fig:attrito}
\centerline{\includegraphics[width=8cm]{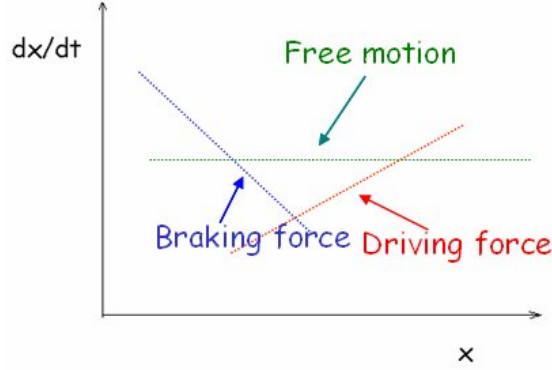}}
\caption{Phase space of a point
particle interacting with an isotropic homogeneous medium. Changing $x$ into
$a,$ the scale factor of a RW universe, the phase space stays unchanged.}
\end{figure}

There is indeed a very simple analogy we may consider. It is synthesised in
fig. (\ref{fig:attrito}). The phase space of a point particle interacting with
an homogeneous isotropic medium is simply bidimensional: the motion of the
particle can only be straight and the relevant parameters are just the
position $x$ and speed $dx/dt$ of the particle. If we now consider a RW
universe we see that its phase space is also bidimensional; it suffices to
change the position of the particle with the scale factor of the universe, $%
a $, and the speed with the rate of change of $a$, $\dot{a}$, and fig (\ref%
{fig:attrito}) is converted into the phase space of the universe: the free
motion corresponds to an inertial expansion (linear increase of $a$); the
effect of a braking force is the equivalent of a decelerated expansion; a
driving force corresponds to accelerated expansion.

Let us then examine the system composed of a point particle and an
homogeneous isotropic medium. There exists a simple situation in which
viscous motion can be described by means of the action integral \cite{caldi}
\cite{mio}:

\begin{equation}
S=\frac{m}{2}\int e^{\left( \alpha t+\beta x\right) /m}\dot{x}^{2}dt
\label{naive}
\end{equation}

The Lagrangian in (\ref{naive}) is rather naive, but it can be recast in a
relativistic invariant form, as%
\begin{equation}
S=m\int e^{\gamma \cdot \gamma }ds=m\int e^{\eta _{\mu \nu }\gamma ^{\mu
}x^{\nu }}ds,  \label{relativo}
\end{equation}%
where the exponent contains the scalar given by the internal product of two
four-vectors, $\gamma =\left( \alpha ,\beta ,\beta ,\beta \right) $ (whose
components are the "viscous" coefficients of the medium) and $r=\left(
t,x,y,z\right) $, which corresponds to the position vector of the particle
in space-time; $ds$ is the line element of the world-line of the particle.
The interaction with the medium is here described by a modification, or, to
say better, an extension of the usual relativistic free-particle Lagrangian.

Exploiting the analogy between the phase spaces, we may conjecture from (\ref%
{relativo}) an action integral for space-time as such \cite{TC}:%
\begin{equation}
S=\int e^{-g_{\mu \nu }\gamma ^{\mu }\gamma ^{\nu }}R\sqrt{-g}d^{4}x.
\label{action}
\end{equation}

The exponent in (\ref{action}) is the simplest scalar we can build combining
the configuration "coordinates" of our manifold (the components of the
metric tensor) with a four-vector. The sign has been chosen with hindsight,
once the interesting consequences of this choice have been worked out. The
rest is the traditional Einstein-Hilbert action. The vector $\gamma $ will
be non-trivial, i.e. a function of the coordinates, whenever defects are
present in the manifold.

\subsection{The accelerated expansion.}

\label{sec:4.1}

We may study the implications of (\ref{action}) in the case of a RW
symmetry, which is induced by a point defect\footnote{%
Actually the same general results hold also in the case that the defect is
represented by any given space-like hypersurface.}. The effective Lagrangian
density is then:%
\begin{equation}
L=e^{-\chi ^{2}}\left( a\ddot{a}+\dot{a}^{2}\right) a,  \label{lagRW}
\end{equation}%
where $\chi $ is the time component of the four-vector $\gamma $ (the only
non-zero component, because of the symmetry). If we introduce for $\gamma $
the typical condition for incompressibility in the elasticity theory, that
would now be $\nabla _{\mu }\gamma ^{\mu }=0$, we obtain%
\begin{equation}
\chi \propto \frac{1}{a^{3}}:  \label{divergenza}
\end{equation}%
a sort of a four-dimensional Coulomb's law.

\begin{figure}\label{fig:quattro}
\centerline{\includegraphics[width=10cm]{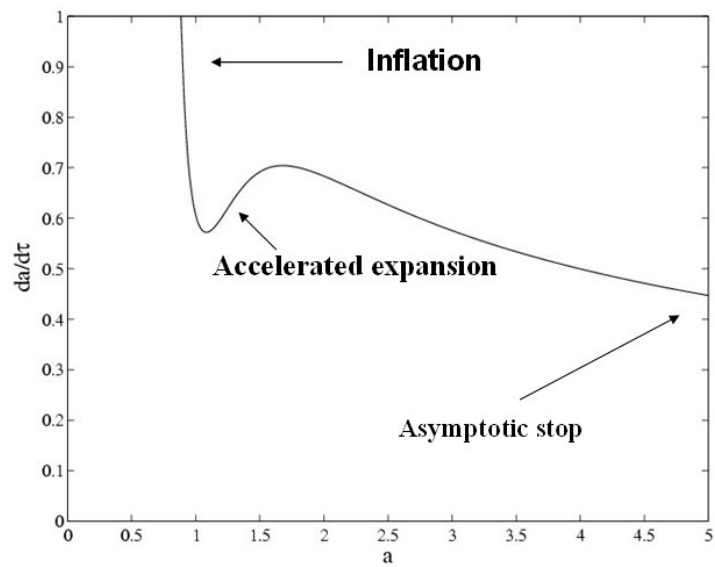}}
\caption{Expansion rate of a RW empty
space-time with a defect in the origin. The behaviour includes an initial
inflationary era, followed by an accelerated expansion era, and finally by a
decreasing expansion rate leading asymptotically to a stop.}
\end{figure}

From (\ref{lagRW}) and (\ref{divergenza}) it is pose to deduce $\dot{a}$ as
a function of $a$, which is shown in fig. (\ref{fig:quattro}) \cite{TC}.
Remarkably, the expansion rate starts from an infinite value, giving rise to
an initial inflationary era; then an accelerated expansion epoch follows;\
finally the expansion starts again to slow down asymptotically reaching zero.

This result is interesting in view of the observed accelerated expansion of
the universe, furthermore displaying a much reassuring asymptotic behaviour,
and giving, as a free gift, also inflation. However it must be reminded that
we are dealing with the empty space time only, so what we are deducing is
the pure effect of the defect at the origin. Matter can be introduced in the
Lagrangian in the usual way, i.e. adding appropriate terms minimally coupled
to the geometry.

This theory, which we have called Cosmic Defect (CD) theory, has indeed been
applied to the fit of the SnIa luminosity data, giving good and encouraging
results \cite{fit}.

\section{The "elastic" approach}

\label{sec:5}

Pursuing literally the elastic analogy we may try another approach to the
definition of the Lagrangian. We may treat the strain induced by the
presence of a defect as a field \emph{in }the space-time rather then a
property \emph{of }the space-time. A weakness of this approach is in the
fact that we already know, for instance, that attempts to treat gravity as a
field in the space-time usually fail. Despite this, let us see how cosmology
would look like. We must consider the elastic energy density associated with
a strained state, which would be $w=\frac{1}{2}\sigma _{\alpha \beta
}\varepsilon ^{\alpha \beta }$. The corresponding action integral for the
(empty) space-time is then:%
\begin{equation}
S=\int \left( R-\frac{\kappa }{2}\sigma _{\mu \nu }\varepsilon ^{\mu \nu
}\right) \sqrt{-g}d^{4}x.  \label{elastico}
\end{equation}

A series of comments regarding the action (\ref{elastico}) are in order.
When introducing a field in a Lagrangian we expect to have both potential
and dynamical terms; the latter are here apparently missing. However we
should remember that the strain tensor, according to (\ref{strain}), does
indeed contain (as well as $\sigma _{\mu \nu }$ does) the displacement
vector $u$ and its first derivatives. The curvature scalar, in turn,
contains up to third order derivatives of $u$, because of (\ref{metric}).
This is why, in a sense, we can interpret $R$ as being the "dynamical" term
in (\ref{elastico}) so that the structure of the Lagrangian is formally the
equivalent of the difference between dynamical and potential terms.

Another remark is that what in (\ref{elastico}) appears to be the usual
minimal coupling between the field (the elastic field) and the geometry is
actually more complicated because the field is also included in the metric
tensor, so that the coupling goes up to higher order terms than with other
fields. Besides this we will also see in a moment that the coupling constant
$\kappa $ is absorbed into other parameters typical of space-time as a
continuum.

Limiting our considerations to the linear elasticity case\footnote{%
An interesting discussion of the non-linear case may be found in \cite%
{unzicker}.} Hooke's law (\ref{hooke}) holds; adding the hypothesis of a
pointlike defect in an homogeneous isotropic space-time, which means RW
symmetry, Hooke's law has the form (\ref{omogeneo}); finally (\ref{elastico}%
) becomes:%
\begin{equation}
S=\int \left( R-\frac{\lambda }{2}\varepsilon ^{2}-\mu \varepsilon _{\alpha
\beta }\varepsilon ^{\alpha \beta }\right) \sqrt{-g}d^{4}x,  \label{esse}
\end{equation}%
with%
\begin{equation*}
\begin{array}{c}
\varepsilon =\frac{\dot{\psi}}{2}\left( 2+\dot{\psi}\right) -\frac{3}{2}%
\frac{\psi ^{2}}{1-\psi ^{2}} \\
\varepsilon _{\mu \nu }\varepsilon ^{\mu \nu }=\frac{\dot{\psi}^{2}}{2}%
\left( 2+\dot{\psi}\right) ^{2}+\frac{3}{4}\frac{\psi ^{4}}{\left( 1-\psi
^{2}\right) ^{2}}.%
\end{array}%
\end{equation*}

As already said, the coupling constant $\kappa $ has been merged with the Lam%
\'{e} coefficients $\lambda $ and $\mu $. Use has been made also of (\ref%
{apsi}).

From (\ref{esse}) the following fourth order equation for $\dot{\psi}$ can
be obtained.%
\begin{equation}
\begin{array}{c}
6\dot{\psi}^{2}\frac{\psi ^{2}}{\sqrt{1-\psi ^{2}}}-\lambda \dot{\psi}\left(
\frac{\dot{\psi}}{2}\left( 2+\dot{\psi}\right) -\frac{3}{2}\frac{\psi ^{2}}{%
1-\psi ^{2}}\right) \left( 1+\dot{\psi}\right) \left( 1-\psi ^{2}\right) ^{%
\frac{3}{2}} \\
-\mu \dot{\psi}^{2}\left( 2+\dot{\psi}\right) \left( 1+\frac{3}{2}\dot{\psi}%
\right) \left( 1-\psi ^{2}\right) ^{\frac{3}{2}} \\
+\frac{\lambda }{2}\left( \frac{\dot{\psi}}{2}\left( 2+\dot{\psi}\right) -%
\frac{3}{2}\frac{\psi ^{2}}{1-\psi ^{2}}\right) ^{2}\left( 1-\psi
^{2}\right) ^{\frac{3}{2}} \\
+\frac{3}{4}\mu \frac{\psi ^{4}}{\left( 1-\psi ^{2}\right) ^{2}}\left(
1-\psi ^{2}\right) ^{\frac{3}{2}}=W.%
\end{array}
\label{cumber}
\end{equation}%
$W$ is a constant.

\subsection{One more analogy}

\label{sec:5.1}

Considering how cumbersome eq. (\ref{cumber}) is, a different approach,
still remaining within the elastic framework, may be envisaged \cite{marsden}%
. It is a simple suggestive analogy. Let us start from eq. (\ref{omogeneo})
in intrinsic coordinates; among the admissible values of the parameters
there is also $\lambda =-\mu $. Suppose this is the case for space-time: the
relation between stress and strain then becomes:%
\begin{equation}
\varepsilon _{\mu \nu }-\frac{1}{2}g_{\mu \nu }\varepsilon =\frac{1}{2\mu }%
\sigma _{\mu \nu }.  \label{epsi}
\end{equation}

Eq. (\ref{epsi}) is formally identical to the Einstein equations for the
gravitational field, so Madsen's \cite{marsden} suggestion is to directly
identify $\varepsilon _{\mu \nu }$ with $R_{\mu \nu }$. Going on along this
play of correspondences, we then use Hooke's law and consider the elastic
potential energy density
\begin{equation}
w=\frac{1}{2}C_{\mu \nu \alpha \beta }\varepsilon ^{\mu \nu }\varepsilon
^{\alpha \beta }=\frac{\lambda }{2}\varepsilon ^{2}+\mu \varepsilon _{\alpha
\beta }\varepsilon ^{\alpha \beta }.  \label{modello}
\end{equation}

On the model of (\ref{modello}) we build the space-time potential:
\begin{equation*}
\Phi =\frac{1}{2}R^{2}-\kappa R_{\mu \nu }R^{\mu \nu }
\end{equation*}%
and use it to write the action integral%
\begin{equation*}
\int \left( \frac{1}{2}R^{2}-\kappa R_{\mu \nu }R^{\mu \nu }\right) \sqrt{-g}%
d^{4}x.
\end{equation*}

Forgetting for a moment the slippery way followed to obtain it, the final
result is a second order theory that could be classified as a special case
of an $f\left( R\right) $ theory.

\section{Generalizing the Cosmic Defect theory}

\label{sec:6}

The version of the CD theory I have outlined in sect. (\ref{sec:4}) is
characterized by the presence in the action integral of the exponential
factor $\exp \left( -g_{\mu \nu }\gamma ^{\mu }\gamma ^{\nu }\right) $ which
gives rise to an extremely steep expansion in the very early era of the
universe. On one side that behaviour is even too fast, on the other in the
Lagrangian no explicit evidence of the dynamics of $\gamma $ appears and, in
order to find the functional form of the four-vector, the incompressibility
condition $\nabla _{\alpha }\gamma ^{\alpha }=0$ has been introduced.

A possible generalization of the action (\ref{action}), which allows to
partially release the initial rather strict conditions, is the following:%
\begin{equation}
S=\int e^{-\gamma _{\mu }\gamma ^{\nu }}R_{\nu }^{\mu }\sqrt{-g}d^{4}x.
\label{azione}
\end{equation}

The exponential term has now to be interpreted in the operatorial sense, and
indeed it is the starting point for a series development:%
\begin{equation}
e^{-\gamma _{\mu }\gamma ^{\nu }}R_{\nu }^{\mu }=R-\gamma ^{\alpha }\gamma
^{\beta }R_{\alpha \beta }+...  \label{sviluppo}
\end{equation}%
When no defect is present we have $\gamma \equiv 0$ and the usual
Hilbert-Einstein action. A defect implies other terms to come into the
arena. We should also remember that, due to the properties of the Riemann
tensor and to the fact that $\gamma $ is a four-vector,%
\begin{equation}
\gamma ^{\beta }R_{\alpha \beta }=\left( \nabla _{\alpha }\nabla _{\nu
}-\nabla _{\nu }\nabla _{\alpha }\right) \gamma ^{\nu }.  \label{rieman}
\end{equation}

Stopping the development in (\ref{sviluppo}) to the first non trivial term
and taking into account (\ref{rieman}) we have the effective Lagrangian:%
\begin{equation*}
L=\left[ R-\nabla _{\nu }\gamma ^{\alpha }\nabla _{\alpha }\gamma ^{\nu
}+\left( \nabla _{\beta }\gamma ^{\beta }\right) ^{2}\right] \sqrt{-g}.
\end{equation*}

Introducing the RW symmetry and again using integration by parts in the
action integral, we have the corresponding so called point Lagrangian:%
\begin{equation}
L=6a\dot{a}^{2}\left( 1+\chi ^{2}\right) +6\chi \dot{\chi}a^{2}\dot{a}.
\label{lagpoin}
\end{equation}%
From (\ref{lagpoin}) the two equations follow:%
\begin{equation}
\begin{array}{c}
\dot{a}^{2}=\frac{W}{a\left( 1+\chi ^{2}\right) } \\
\chi a^{2}\ddot{a}=0.%
\end{array}
\label{equazioni}
\end{equation}

The solution of (\ref{equazioni}) is $\dot{a}=V=$ constant ($W$ too is a
constant) with $\chi =\sqrt{W/\left( V^{2}a\right) -1}$. So this model
describes a uniformly expanding space-time. The structure contained in the
CD theory (see fig. (\ref{fig:quattro})) has disappeared, after adopting (\ref%
{azione}) and the limited development (\ref{sviluppo}).

As a matter of fact (\ref{azione}) does not contain (\ref{action}) as a
special case. If we wish a real generalization of CD we can use a Lagrangian
density like:%
\begin{equation}
e^{-\left( \delta _{\alpha }^{\nu }\delta _{\mu }^{\beta }+\delta _{\alpha
}^{\beta }\delta _{\mu }^{\nu }\right) \gamma ^{\alpha }\gamma _{\beta
}}R_{\nu }^{\mu }\sqrt{-g},  \label{ampio}
\end{equation}%
which includes the one in (\ref{action}).

Instead of (\ref{sviluppo}) we now have:%
\begin{equation*}
e^{-\left( \delta _{\alpha }^{\nu }\delta _{\mu }^{\beta }+\delta _{\alpha
}^{\beta }\delta _{\mu }^{\nu }\right) \gamma ^{\alpha }\gamma _{\beta
}}R_{\nu }^{\mu }=R\left( 1-\gamma _{\mu }\gamma ^{\mu }\right) -\gamma
^{\alpha }\gamma ^{\beta }R_{\alpha \beta }+...
\end{equation*}

The consequent effective point Lagrangian density (RW symmetry) is:%
\begin{equation*}
L=6a\dot{a}^{2}-6\chi \dot{\chi}a^{2}\dot{a},
\end{equation*}%
however the situation does not really change in the sense that one obtains
an expansion $\propto \tau ^{2/3}$, typical of a Friedman-Robertson-Walker
matter dominated universe, independent from $\chi $.

Evidently the properties of the CD model are contained in the higher order
terms of (\ref{ampio}).

Of course there are many ways in which one can further generalize the ansatz
(\ref{ampio}). One can for instance introduce a "potential" term $\gamma
^{2}=\gamma _{\alpha }\gamma ^{\alpha }$ in the Lagrangian, considering
that, notwithstanding its geometric interpretation, $\gamma $ is anyway a
vector field and its energy content must directly influence curvature. One
could also parametrize the Lagrangian, writing:%
\begin{equation*}
S=\int \left[ R\left( 1+\sigma \gamma ^{2}\right) +\lambda \nabla _{\alpha
}\gamma ^{\beta }\nabla _{\beta }\gamma ^{\alpha }+\mu \left( \nabla _{\mu
}\gamma ^{\mu }\right) ^{2}+\nu \nabla _{\alpha }\gamma ^{\beta }\nabla
^{\alpha }\gamma _{\beta }\right] \sqrt{-g}d^{4}x,
\end{equation*}%
however, following the thread of conjectures one looses more and more the
contact with the, though fragile, initial physical motivation.

\section{Conclusion}

\label{sec:7}

We have reviewed here an approach to the description of space-time based on
the elastic continuum analogy, integrated with the possible presence of
defects, in the sense of Volterra's description \cite{volterra}. Once this
scheme is adopted we have seen that the approach is not unique. Among other
possibility I have privileged the named Cosmic Defect theory, which proved,
simultaneously, to be manageable and to give good results when trying to
describe the accelerated expansion of the universe \cite{fit}.

Actually regarding space-time at the cosmological level a real "forest" of
theories exists, mostly based either on the concept of dark energy (from
cosmological constant \cite{cosmol} to phantom energy \cite{darkreview}), or
on modifications or extensions of General Relativity (from MOND \cite{mond}
to $f(R)$ theories \cite{freview}); not considering quantum theories
(strings \cite{maartens} or loop quantum gravity \cite{lqg}). Most often
these numerous approaches belong to what I would call "Lagrangian
engineering": let us somehow change the Lagrangian and see what happens.
These attempts can be more or less fortunate and more or less \emph{ad hoc}%
, but generally rely on rather staggering physical bases, looking for
\emph{ex post} justification. It is also often possible to see that
apparently different theories and approaches are indeed related to each
other and lead, totally or partly, to convergent or coincident results. For
example the whole elastic analogy approach is formally a (group of)
vector-tensor theory, being based on the displacement vector field and
related strain tensor field. We verified that the CD theory also is
reducible to a special case of general vector theories \cite{nrat}; Madsen's
conjecture described in sec. (\ref{sec:5.1}) leads to a sort of second
degree $f(R)$ Lagrangian. Furthermore conformal transformations can convert
modified or extended gravity theories into GR plus some more or less exotic
dark energy fluid (CD is again an example).

In this very jungle I think it is better to have a compass pointing in some
direction, rather than moving around blindly in pursuit of a local and
ephemeral success. In other words it is better to start from some physical
paradigm that suggests where to go and what to look for. This is why, thanks
also to the initial positive results, I think the elastic continuum model
and the CD theory are a good conceptual framework that deserves further
exploration and deepening.

%\printindex

\end{document}